\UseRawInputEncoding
\documentclass[%
 reprint,
 amsmath,amssymb,
 aps,
]{revtex4-1}

\usepackage{graphicx}
\usepackage{bm}
\usepackage{hyperref}
\usepackage{braket}
\usepackage{commath}
\usepackage{color}
\usepackage{subfigure}
\usepackage{epstopdf}
\usepackage{epsfig}
\usepackage{amsmath}
\usepackage{mathtools}
\usepackage{dcolumn}
\usepackage{url}
\usepackage{hyperref}
\hypersetup{%
        plainpages=true,
        breaklinks=true,
        hypertexnames=false,
        pageanchor=true,
        colorlinks=true,
        linkcolor={blue},
        citecolor={magenta},
        urlcolor={blue},
        anchorcolor={black}
      }

\usepackage{mathrsfs,bm,amsthm,amsfonts,xspace,float,bookmark}
\usepackage{textcomp}
\usepackage{tabu}
\usepackage{bbold}
\usepackage{xspace}
\usepackage{float}
\usepackage[inline]{enumitem}

\makeatletter
\newcommand*{\closeindex}[1]{_{\mkern-4.5mu#1}}
\DeclareRobustCommand{\power}{\textrm{\textit{\textsf{P}}}\@ifnextchar_{\expandafter\closeindex\@gobble}{}}
\DeclareRobustCommand{\gain}{\textrm{\textit{\textsf{G}}}\@ifnextchar_{\expandafter\closeindex\@gobble}{}}
\makeatother

\newcommand{\unite}[1]{\,\rm#1}
\newcommand{\Gama}[1]{{\it{\Gamma_{\rm#1}}}}
\newcommand{\Omga}[1]{{\it{\Omega_{\rm#1}}}}

\newcommand{\Gamaupdown}[2]{{\it{\Gamma_{\rm#1}^{\rm#2}}}}

\usepackage{amsmath}
\DeclareMathOperator{\Tr}{Tr}

\begin{document}

\title{Propagating Wigner-negative states generated from the steady-state emission of a superconducting qubit}

\author{Yong Lu}
\email[e-mail:]{yongl@chalmers.se}
\author{Ingrid Strandberg}
\author{Fernando Quijandr\'{\i}a}
\author{G\"{o}ran Johansson}
\email[e-mail:]{goran.l.johansson@chalmers.se }
\author{Simone Gasparinetti}
\author{Per Delsing}
\email[e-mail:]{per.delsing@chalmers.se}
\affiliation{Department of Microtechnology and Nanoscience MC2, Chalmers University of Technology, SE-412 96
G\"oteborg, Sweden}

\date{\today}%

\begin{abstract}
We experimentally demonstrate the steady-state generation of propagating Wigner-negative states from a continuously driven superconducting qubit. We reconstruct the Wigner function of the radiation emitted into propagating modes defined by their temporal envelopes, using digital filtering. For an optimized temporal filter,
we observe a large Wigner logarithmic negativity, in excess of 0.08, in agreement with theory. The fidelity between the theoretical predictions and the states generated experimentally is up to $99\%$, reaching state-of-the-art realizations in the microwave frequency domain. Our results provide a new way to generate and control nonclassical states, and may enable promising applications such as quantum networks and quantum computation based on waveguide quantum electrodynamics.
\end{abstract}

\maketitle





In the continuous-variable (CV) approach to quantum information processing~\cite{menicucci2014fault,gu2009quantum}, information can be represented by a phase space quasiprobability distribution such as the Wigner function~\citep{Cahill1969Jan}.
To obtain a quantum computational advantage with a CV quantum circuit, either the input state, the circuit itself, or the final measurement needs to be characterized by a negative quasiprobability distribution~\citep{Rahimi-Keshari2016Jun}. For example, Gaussian boson sampling~\citep{Hamilton2017Oct, Zhong2020Dec} utilizes Wigner-positive input states and a linear circuit, but has photon-number resolving detectors which are associated with Wigner-function negativity~\citep{Hlousek2021Jan}.
%
In this work, we focus on generating Wigner-negative states that could be used as a computational resource in a linear circuit with heterodyne or homodyne measurements.

Nonclassical states with negative Wigner functions have been implemented using natural atoms~\cite{hacker2019deterministic,deleglise2008reconstruction}, trapped ions~\cite{monroe1996schrodinger, Fluhmann2019Feb} and optical photons~\cite{makino2016synchronization}. In superconducting quantum circuits, some Wigner-negative states have been implemented in 3D cavities~\cite{campagne2020quantum, leghtas2015confining,wang2016schrodinger,campagne2020quantum} and resonators~\cite{hofheinz2008generation,chu2018creation,satzinger2018quantum} where the states are stored in the confined modes of the cavities. Such states have a limited lifetime and the corresponding setups are relatively complicated. In waveguide quantum electrodynamics by using superconducting circuits, single photons as well as cat states have been generated and transferred~\cite{peng2016tuneable,houck2007generating,pfaff2017controlled}. These propagating states are generated by either releasing a cavity state or by exciting a quantum emitter that subsequently decays
into a waveguide. However, in both cases, the states result from transient dynamics so that resetting the system after a certain time is necessary. A continuously driven source may lead to higher generation rates, but the question whether Wigner-negative states can be generated this way has been theoretically addressed only recently~\cite{Strandberg2019Dec,Quijandrna2018Dec}, and experimental verification was still lacking. In Ref.~\cite{hoi2011demonstration} a qubit in an \emph{infinite} waveguide was used to demonstrate photon antibunching or nonclassical correlations between photons emitted by the qubit.  While that is a nonclassical phenomenon, the Wigner function  of any propagating mode would be positive since the emitted field is in a mixed state \cite{Strandberg2019Dec}.

\begin{figure}[tbph]
\includegraphics[width=1\linewidth]{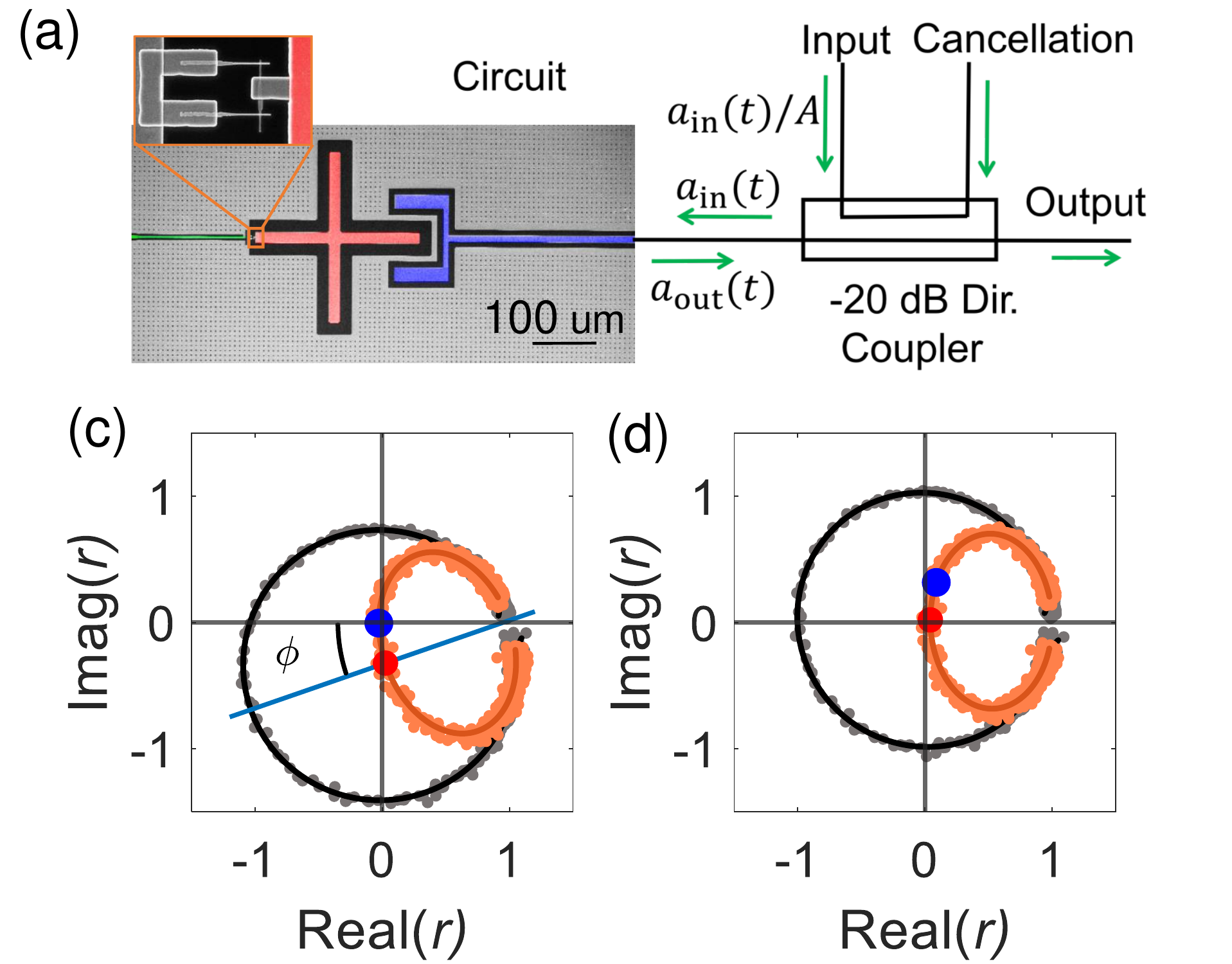}
\caption{\textsf{{\bf{\textsf{Measurement setup and spectroscopy of a transmon qubit.}}}
{(a)} A simplified schematic of the measurement setup. {Dir.Coupler denotes a directional coupler with coupling factor $A=0.1$. $a_{\rm{in}}(t)$ and $a_{\rm{out}}(t)$ are the input and output signals. A superconducting circuit is connected to the directional coupler, where a transmon consisting of a superconducting island (red) shunted by a SQUID (Superconducting QUantum  Interference  Device) loop is capacitively coupled to a coplanar waveguide (blue) and inductively coupled to a flux line (green). The inset shows a close-up of the SQUID loop.}
{(b)} Single-tone spectroscopy, showing real and imaginary parts of the reflection coefficient \textsf{\textit{r}} of a probe with the qubit at zero external flux. The black and orange dots are the experimental data of the reflection coefficient at two different intensities of the probe, namely, $\Omga{}\ll\Gama{r}$ and $\Omga{}\approx 0.707\Gama{r}$, respectively. The solid curves show the corresponding fittings to theory.
{(c)} Here, the data in {\bf{c}}} has been corrected for the impedance mismatch in the probe line \cite{suppl}.
}
\label{setup}
\end{figure}

In this work, we experimentally demonstrate the generation of Wigner-negative states using the steady-state emission from a continuously driven superconducting qubit. We study the nonclassical properties of the quantum state of light propagating along a \emph{semi-infinite} waveguide, occupying the single mode defined by a temporal filter. We reconstruct the Wigner function of the state, and investigate the effect of different temporal mode filters.  Our flux-tunable transmon qubit \cite{koch2007charge} is capacitively coupled to the open end of a one-dimensional transmission line [Fig.~\ref{setup}(a)and (b)]. The circuit is equivalent to an atom in front of a mirror in 1D space with a negligible distance between the qubit and the mirror. In this work, the qubit is operated at zero external flux.
%
%
%

The total scattered field from the qubit in front of a mirror is characterized by its annihilation operator $a_{\rm{out}}$ which contains two contributions: the incoming field operator $a_{\rm{in}}$ reflected by the mirror and the field emitted by the qubit, according to~\citep{Yurke1984Mar, Gardiner1985Jun}

\begin{equation}
a_{\rm{out}}(t) = a_{\rm{in}}(t) - i\sqrt{\Gama{r}e^{i\phi}} \sigma_-(t),
\label{output0}
\end{equation}
where $a_{\rm{in}}(t)={\Omga{}}(t)/({2\sqrt{\Gama{r}e^{i\phi}}})$, $\Gama{r}$ is the decay rate of the qubit into the transmission line, $\Omga{}(t)$ is the Rabi frequency of the coherent input, the phase $\phi$ quantifies the impedance mismatch in the line (for a perfectly matched line, $\phi=0$; see~\cite{suppl}), and
 $\sigma_-(t)$ is the qubit lowering operator.

To characterize the device, {we apply a coherent continuous probe to the input port in Fig.~\ref{setup}(a). An input signal $a_{\rm{in}}(t)$ reaches the qubit. The scattered field $a_{\rm{out}}(t)$ is measured at the output port. In the steady state, the expectation value of $\langle {a_{\rm{out}}(t)}\rangle$ is constant. By taking averages, we obtain the complex reflection coefficient as $r = \langle {a_{\rm{out}}}\rangle /\langle{a_{\rm{in}}}\rangle$}
[Fig.~\ref{setup}(b), black dots]. By fitting data to the theory [Fig.~\ref{setup}(b), black line], we obtain $\phi=-0.319\pm0.03$, $\Gama{r}/2\pi=1.11\unite{MHz}$ and $\Gama{2}/2\pi=0.528\unite{MHz}$, where $\Gama{2}$ is the total decoherence rate of the qubit.
After compensating for the impedance mismatch, a parametric plot of the reflection coefficient describes a circle in the IQ plane [Fig.~\ref{setup}(c)].
The radius of the circle approaches unity, which implies than nonradiative decay and pure dephasing are negligible in our sample.




For a coherently driven two-level system in our mirror-like geometry, the largest Wigner negativity is expected when driving on resonance and choosing the drive power so that the coherent reflection vanishes~\cite{Quijandrna2018Dec}. This effect, reported in previous experiments~\cite{hoi2015probing}, is due to destructive interference between the radiation reflected by the mirror and the radiation coherently scattered by the two-level system. Here, we also find that at this drive power the coherence between the qubit's ground and excited states is maximized~\cite{suppl}.  We call this power the critical power. In our case, due to the impedance mismatch, full cancellation is achieved at a slightly detuned driving frequency [by $170\unite{kHz}$, blue circle in Fig.~1(b)]. By contrast, resonant driving of the system at the same power gives some residual coherent reflection (red circle in Fig.~1(b), caused by the impedance mismatch), which we eliminate in the measurements below by applying a cancellation pulse to the cancellation port of the directional coupler [Fig.~\ref{setup}(a)]. As we show in the following, both on- and off-resonant cases give comparable results with regards to the observed Wigner negativity.


\begin{figure}[t]
\includegraphics[width=0.8\linewidth]{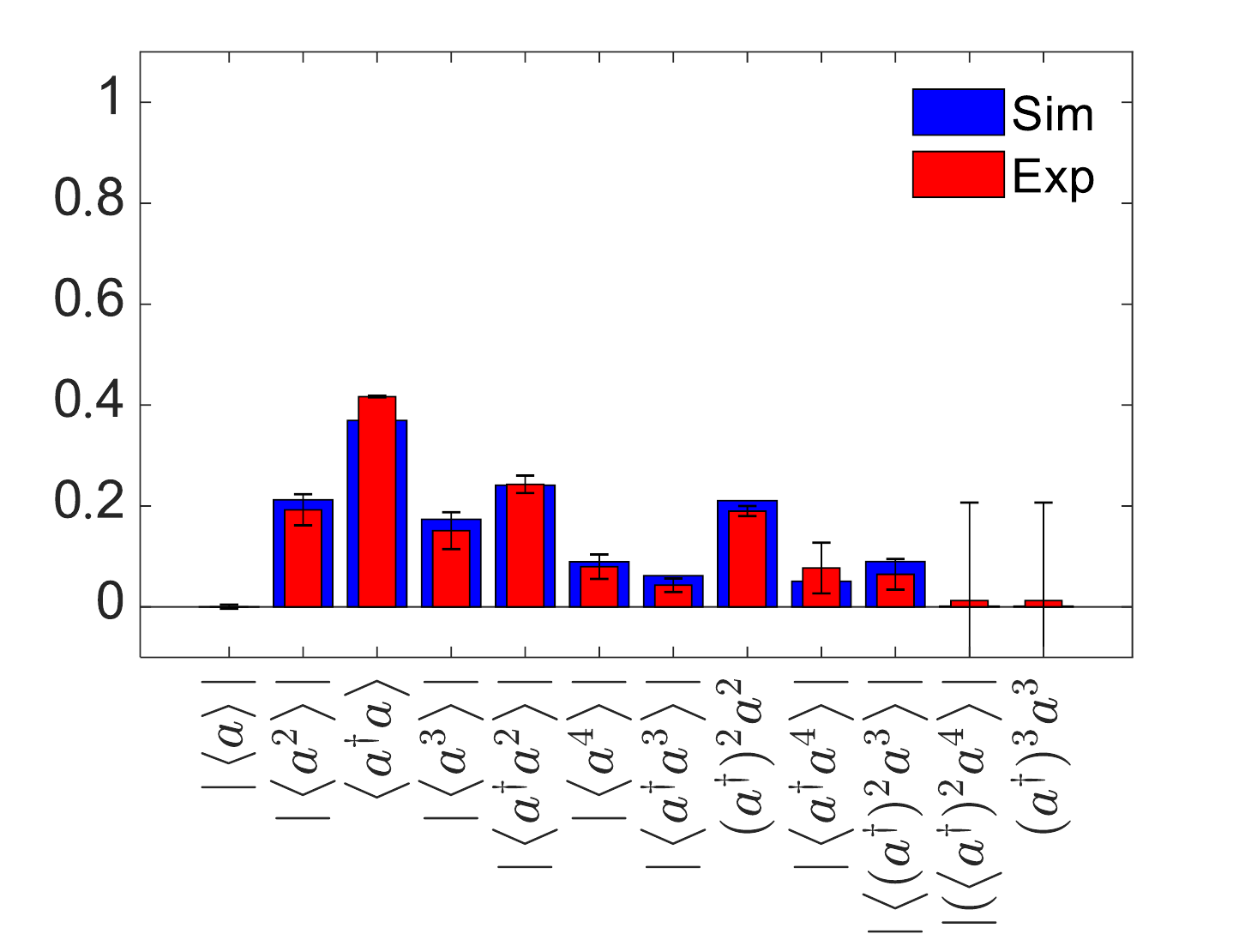}
\caption{\textsf{{\bf{\textsf{{Moments for the on-resonance case.}}}}
Different orders of moments for the propagating state from the qubit emission with the measurement time window $\tau=\sf{2.0}$, where $\tau$ defines the length of the boxcar filter in the time domain. The red rectangles correspond to experimental data, which includes the impedance mismatch in the line. The blue rectangles corresponds to a numerical simulation of the ideal line, without the impedance mismatch.
{
}
}}
\label{moment}
\end{figure}

\begin{figure*}[tbph]
\includegraphics[width=0.9\linewidth]{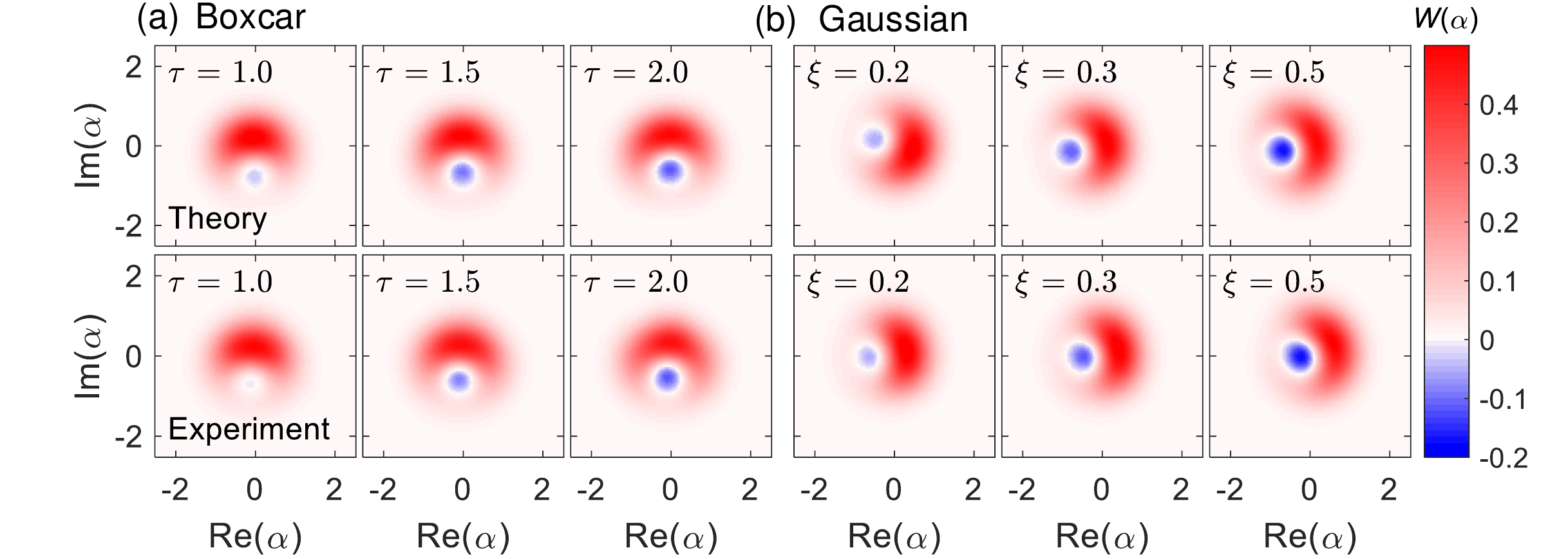}
\caption{\textsf{{\bf{\textsf{{Reconstructed Wigner function with boxcar and Gaussian filters.}}}} Comparison between the numerical simulation and the experiment for Wigner functions of the propagating states.
{Wigner functions for: (a) the on-resonance case using a boxcar filter of length $\tau/\Gama{2}$.
(b) the off-resonance case using a Gaussian filter of standard deviation $\xi/\Gama{2}$. The negative values indicate that the states are nonclassical. The color scale has been adjusted to the measured data range for optimal visibility. The rotation between (a) and (b) is due to the phase of the driving field.}}
}
\label{wignerpic}
\end{figure*}

{The emission from the qubit is not confined and comprises a continuum of modes described by the operator $a_{\rm{out}}(t)$. Here we study the quantum state occupying the single mode $a$ defined by the filter function $f(t)$. Consequently, the properties of the state will depend on the chosen $f(t)$.}
We perform a full tomography of propagating modes, by measuring their statistical moments $\langle a^{\dagger m} a^n \rangle ,\,\, m,n = 0, \dots, N_c$, where $a= \int_0^\infty {\rm d}t\, f(t) a_{\rm{out}}(t)$ \cite{eichler2012characterizing} and $N_c$ is the photon-number cutoff. In order to excite the qubit, we send a long pulse $4.4\unite{\mu s}$, much longer than the lifetime of the qubit, $T_1=1/\Gama{1}=145\unite{ns}$. This in order to ensure that the qubit has reached the steady-state before we perform the measurement of its emission field~\cite{suppl}. First, we consider a normalized rectangular boxcar filter which is a constant function within the time interval from $t_0=2.5~\mu s$ to $t_0+\tau/\Gama{2}$ and zero elsewhere. In Figure.~\ref{moment} we show the magnitude of the moments up to 6th order ($m+n\le 6$) for $\tau=2$  for the on-resonant case in red rectangles.
Due to the small average photon number $\langle a^\dagger a \rangle$, it is enough to truncate
the Fock space up to $N_c = 4$. Our results are very similar to the numerical simulations \cite{johansson2012qutip}, using the method presented in \cite{Kiilerich2019Sep} which correspond to the blue rectangles. The slight difference is due to the impedance mismatch which gives a higher Rabi frequency $\Omga{m}=(0.74\pm0.01)\Gama{1}$~\cite{suppl} than the ideal case  $\Omga{m}= 0.707\Gama{1}$, leading to a slightly larger photon number  $\langle a^{\dagger} a\rangle$. Finally, as expected, the first order moment $\langle a \rangle$ is almost zero due to the additional pulse which corrects the effect from the impedance mismatch.

To demonstrate the nonclassical character of the generated states we now turn to their Wigner functions. We first extract the density matrix $\rho$ of the filtered output from the measured moments using maximum likelihood estimation \cite{eichler2011experimental}. Then, we obtain the Wigner function from the relation $W(\alpha)=(2/{\pi})\Tr{[\hat{D}(\alpha)\rho \hat{D}^{\dag}(\alpha)\hat{\Pi}]}$, where $\hat{D}(\alpha)$ is the displacement operator with amplitude $\alpha$ and $\hat{\Pi}$ is the parity operator \cite{eichler2012characterizing}. In Fig.~\ref{wignerpic}(a), the nonclassical nature of the outgoing field is corroborated by the negative values of the Wigner functions for different values of $\tau$. As seen in the plots, with increasing $\tau$ from $0.5$ to $2.0$, the negativity region becomes larger. For $\tau>2.0$, the corresponding negativity is decreased, as will be discussed below.

To quantify the nonclassical content of the state, we use the Wigner logarithmic negativity \cite{albarelli2018resource} defined as ${{\sf{WLN}}=\log \left(\int {\rm d}\alpha\, \vert W(\alpha)\vert \right)}$, which has the property $\sf{WLN} > 0$ when the Wigner function, $W(\alpha)$, has a negative part.

\begin{figure*}
\includegraphics[width=0.8\linewidth]{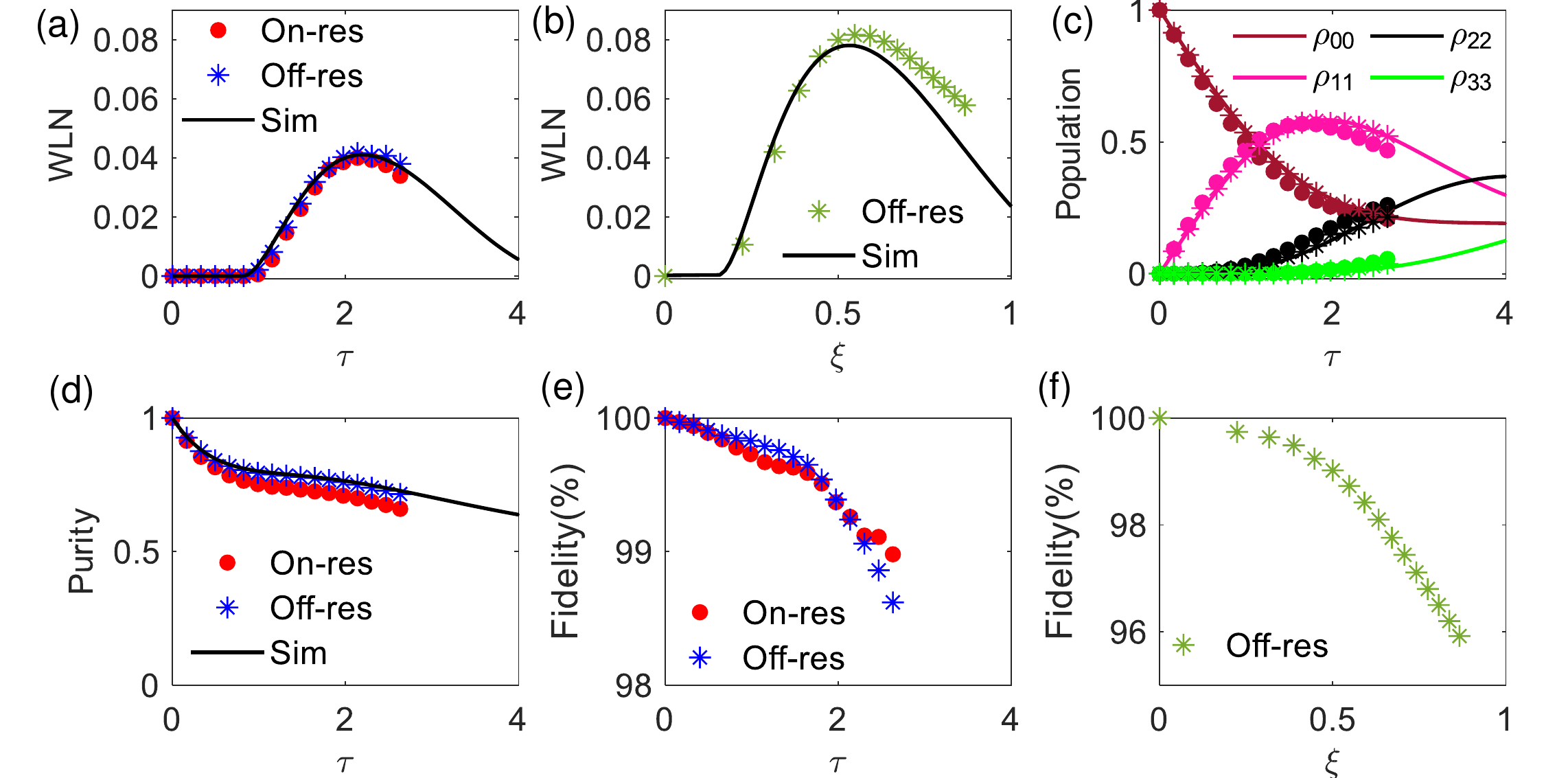} 
\caption{\textsf{{\bf{\textsf{{Wigner logarithmic negativity (WLN), photon number content and purity with different filters.}}}}
These are obtained from the reconstructed density matrix $\rho$ of the filtered output field using maximum likelihood estimation. In all panels, the makers are from the experiment whereas the solid curves are from the numerical simulation. {On/Off-res and Sim represent the on/off resonant cases and the numerical simulation, respectively.}
{(a)} WLN with different $\tau$ values for a boxcar filter. The red and blue markers are for the on- and off-resonant cases, respectively.
{(b)} WLN with different $\xi$ values of the standard deviation of a Gaussian filter.
{(c)} Photon populations for data shown in {\bf{a}}. $\rho_{\sf{nn}}$ are the diagonal elements of the density matrix and thus present the population of the photon number n of the emission field.
{(d)} Photon purity for data shown in {\bf{a}}. The purity is given by $Tr(\rho^2)$.
{(e) and (f) are the corresponding fidelities between experimental and numerical results for the states used in (a) and (b), respectively. The fidelities are calculated according to $F=\Tr{\sqrt{\rho^{1/2}\rho_{\rm{sim}}\rho^{1/2}}}$ where $\rho_{\rm{sim}}$ is the numerical density matrix.}
}
}
\label{wln}
\end{figure*}

We calculate the $\sf{WLN}$ for both on-and off-resonant cases for the boxcar filter
(red and blue markers respectively) in Fig.~\ref{wln}(a). We see that the $\sf{WLN}$ increases when $\tau$ is increased from 0 to $2.1$ whereas it starts to decrease when $\tau$ is increased further. {Our experimental results (markers in Fig.~\ref{wln}) are in excellent agreement with numerical simulations (solid lines)~\cite{kiilerich2019input}.}
Since the chosen filter function defines the observed mode it is reasonable to expect that a different choice of filter function will also affect the observed negativity of the Wigner function [Fig.~\ref{wignerpic}(a) and (b)]. In Fig.~\ref{wln}(b), we show the values of $\sf{WLN}$ for a measurement with a normalized Gaussian filter $f(t)=\sqrt{\Gama{2}}\exp(-t^2\Gamaupdown{2}{2}/4\xi^2)/(2\pi\xi^2)^{1/4}$.
{Compared to the boxcar filter, the Gaussian filter with a width given by $\xi=0.5$ can produce a state with a maximum $\sf{WLN}$ twice as large, {which is also seen by comparing Fig.~\ref{wignerpic}(b) and Fig.~\ref{wignerpic}(a).}
Through numerical optimization we have verified that the Gaussian filter is indeed the optimal filter for maximizing the Wigner negativity, the same method is also used in \cite{strandberg2020wigner}.}

From Eq.~(\ref{output0}), the effect of the driving field on the total output is to displace the emission from the qubit. A displacement operation in phase space amounts to a translation of the Wigner function which does not affect its negativity. Therefore, in order to obtain the field emitted by the qubit, we can remove the coherent signal from the drive by means of a digital displacement with the opposite sign on the extracted density matrix of the total emission. Since we know the Rabi frequency $\Omga{m}$ at the critical power, we can obtain the displacement as $(\Omga{m}/2\sqrt{\Gama{r}/\Gama{2}})\sqrt{\tau}$. Therefore, we can obtain the photon number from the qubit emission in Fig.~\ref{wln}(c). {This figure reveals the relation between the $\sf{WLN}$ and the single-photon population.
{For filtering times much  smaller than  the qubit decay time ($\tau \ll 1$), the field is approximately in the vacuum state, i.e., a Gaussian state and consequently, a Wigner positive state. As we increase $\tau$, the single-photon population becomes non-negligible. In fact, for $0 \leq \tau \leq 1$, the state is mostly a superposition of the vacuum and single-photon states. As shown in Refs.~\cite{Quijandrna2018Dec,Strandberg2019Dec}, in this two-dimensional space, the relation between Wigner negativity and single-photon content strongly depends on the purity. In simple terms, for a statistical mixture of vacuum and a single photon, we require an average population of at least half a photon in order for the state to be Wigner negative. This required population decreases with an increasing purity. By comparing Figs.~\ref{wln}(a) and (c), we see that the $\sf{WLN}$ becomes nonzero roughly when the vacuum and the single-photon populations become identical at $\tau \simeq 1$.}

{For $\tau \geq 1$, two- and three-photon states in Fig.~\ref{wln}(c) are no longer negligible. Nevertheless, the single-photon population becomes the dominant contribution to the state. In fact, the $\sf{WLN}$ achieves its largest value when the single-photon state achieves its largest population
($\rho_{11} \approx 2 \rho_{00}$ and $\rho_{11} \approx 4\rho_{22}$) at $\tau \simeq 2$.}

{Increasing $\tau$ further beyond the qubit decay time, the output field will contain more uncorrelated emissions. As it can be seen in Fig.~\ref{wln}(d), the net effect of this is to reduce the purity of the output state and consequently the Wigner negativity.}

Due to the impedance mismatch, the Rabi frequency for the on-resonant case is slightly higher than the off-resonant case, resulting in a lower purity compared to the off-resonant case [Fig.~\ref{wln}(d)]. Accordingly, the corresponding $\sf{WLN}$ is smaller [Fig.~\ref{wln}(a)].

The single photon population and the purity of the state are higher for a Gaussian than for a boxcar filter, leading to a higher negativity [Fig.~\ref{wln}(a,b) and \cite{suppl}]. {With a larger filter width, the photon population is increased, this will lead to a non-negligible contribution from higher order moments. The measurement of these is limited by the system noise. Nevertheless, lower order moments still yield the dominant contribution to the field state. Therefore, the observed tendency of the $\sf{WLN}$ for $\xi > 0.5$ in Fig.~\ref{wln}(b) is in agreement with the numerical simulations.}

{Our nonclassical states have above $95\%$ fidelities between the experimentally produced states and the predicted states for both types of filters with different lengths [Fig.~\ref{wln}(e) and (f)]. Especially, at maximum values of $\sf{WLN}$, the fidelities are $99.24\%$ and $99.05\%$ for the boxcar and Gaussian filters, respectively.}

Numerically, we find that our non-classical state is much more sensitive to pure dephasing compared to non-radiative decay. Such a non-classical state with biased noise may be useful for quantum computation \cite{aliferis2008fault}. Even though we do not investigate the frequency tunability of our non-classical source by measuring the Wigner function at different qubit frequencies, it is still possible to estimate the pure dephasing and the non-radiative decay rates, and how much they affect the negativity. Our evaluation shows that the tunable bandwidth of our non-classical source can be up to 400\unite{MHz} with negativities above 0.04 using a Gaussian filter~\cite{suppl}.

{Our setup provides a straightforward way to generate nonclassical states. Compared to pulsed operation ~\cite{hofheinz2008generation,leghtas2015confining,wang2016schrodinger,chu2018creation,satzinger2018quantum,campagne2020quantum,pfaff2017controlled,besse2020parity,peng2016tuneable,houck2007generating}, it has several advantages. i) It has a higher generation rate since the pulsed case requires to wait until the qubit returns to the ground state. ii) We operate at the critical power, generating states with unit efficiency whereas in the pulsed scheme many photons are needed for the excitation pulse~\cite{peng2016tuneable,houck2007generating}. iii) For a pulsed source there is a trade-off between high quantum efficiency and avoiding population of the higher levels of the qubit. In our case, the high-level excitation is negligible $\approx 10^{-5}$. iv) Using our source in a quantum network, there is no timing requirement, the receiver can select any time slot of the continuous stream.}

Our experimental results demonstrate that nonclassical states useful for quantum computation can be obtained from the steady-state dynamics of a continuously driven quantum system by applying optimized filters to its propagating output field. These conclusions can be extended to a variety of physical systems~\cite{laucht2012waveguide,katsumi2018transfer}. Recent theoretical work indicates that propagating Wigner-negative states may also be obtained from driven systems whose steady-state intracavity field is Wigner-positive, such as Kerr parametric oscillators \cite{strandberg2020wigner}. This finding further broadens the class of systems to which the techniques shown here are applicable.

We acknowledge the use of Nano-fabrication Laboratory (NFL) at Chalmers. We also acknowledge IARPA and Lincoln Labs for providing the TWPA used in this experiment.
We wish to express our gratitude to Lars J\"{o}nsson, Andreas Bengtsson and Daniel Perez Lozano for help.
This work was supported by the Knut and Alice Wallenberg Foundation via the Wallenberg Center for Quantum Technology (WACQT) and by the Swedish Research Council.




\end{document}


\title{Supplementary Material for "Propagating Wigner-negative states generated from the steady-state emission of a superconducting qubit"}

\author{Yong Lu}
\email[e-mail:]{yongl@chalmers.se}
\author{Ingrid Strandberg}
\author{Fernando Quijandr\'{\i}a}
\author{G\"{o}ran Johansson}
\email[e-mail:]{goran.l.johansson@chalmers.se }
\author{Simone Gasparinetti}
\author{Per Delsing}
\email[e-mail:]{per.delsing@chalmers.se}
\affiliation{Department of Microtechnology and Nanoscience MC2, Chalmers University of Technology, SE-412 96
G\"oteborg, Sweden}

\maketitle
\date{\today}%

\renewcommand{\thefigure}{S\arabic{figure}}
\renewcommand{\thesection}{S\arabic{section}}
\renewcommand{\theequation}{S\arabic{equation}}

\onecolumngrid

\setcounter{section}{0}
\setcounter{equation}{0}

\setcounter{figure}{0}
\section{Measurement setup}\label{sec:supsetup}
\begin{figure*}[tbph]
\includegraphics[width=1\linewidth]{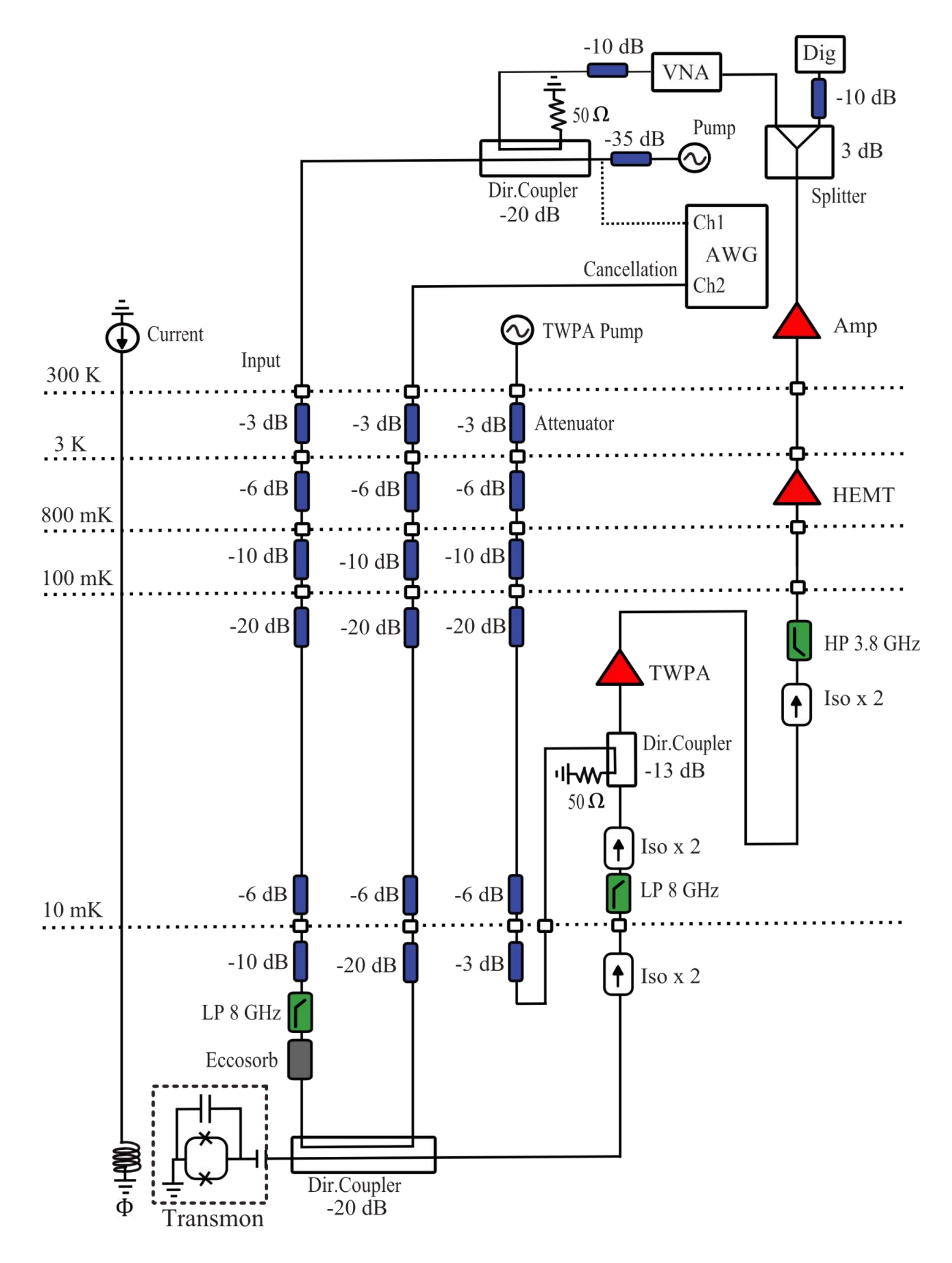}
\caption{\textsf{{\bf{\textsf{{Measurement setup.}}}}
VNA, Iso, LP and HP,TWPA, HEMT and Amp denote a vector network analyzer, an isolator, a low-pass and high-pass filter, a traveling-wave parametric amplifier,  a high electron
mobility transistor amplifier and room temperature amplifiers, respectively. {The blue rectangular boxes are the attenuators with specific attenuations indicated in the figure. The current is sent to the on-chip flux line to generate the external magnetic flux $\Phi$ to tune the qubit frequency. }
}
}
\label{CompleteSetup}
\end{figure*}

Fig. \ref{CompleteSetup} shows our measurement setup where a transmon qubit is weakly coupled to a 1D semi-infinite transmission line. An arbitrary waveform generator (AWG) shapes  the waveform of input coherent photons, and a digitizer captures the output signal, enabling us to measure this propagating mode in the time domain. A vector network analyzer (VNA) is used to measure the reflection coefficient. {In details, to measure the reflection coefficient, we turn off the ouput of AWG and the pump which can generate coherent continuous microwaves, and then turn on a coherent probe coming from a VNA in Fig. \ref{CompleteSetup}. This signal, attenuated by 10\unite{dB}, goes into the input line of the low-temperature refrigerator through a 20\unite{dB} directional coupler. After heavy attenuation inside the refrigerator, the signal bypasses another directional coupler at 10\unite{mK} stage and interacts with the qubit through the capacitive coupling. The signal then is reflected by scattering of the qubit. The reflected signal passes through linear, phase preserving amplification chain including a traveling-wave parametric amplifier (TWPA) \cite{macklin2015near}, a high electron
mobility transistor amplifier (HEMT) and room-temperature amplifiers (Amp) before reaching the VNA. By taking the ratio between the signal received by the VNA and its output, and then normalizing the ratio to the background reference by tuning the qubit away from the probe frequency, we obtain the complex reflection coefficient~\cite{wen2019large,lu2019characterizing}. After characterizing the sample with VNA, we turn it off and generate an input (cancellation) pulse from the AWG to feed into the input (cancellation) line to excite the qubit (cancel the residual coherent photons from the input pulse for the on-resonance case). Then, the qubit emission after amplification is captured by a digitizer. By calibrating the gain of the system from the on-resonance Mollow triplet [see section \ref{sec:supgain}], we obtain the emission power from the qubit.  After applying the digital filters to the output signal, we get the different order of moments to reconstruct the density matrix of the output state based on the maximum likelihood method. Thus, the corresponding Wigner function is calculated from the density matrix. In order to measure the qubit fluorescence as shown in Fig.~\ref{gain}(a), we turn off the VNA and AWG and turn on the pump to drive the qubit on resonance. By measuring the emission with the digitizer in the time domain, we can calculate the corresponding power spectrum density \cite{lu2019characterizing}}.
\section{System calibration}\label{sec:supgain}
\begin{figure*}
\includegraphics[width=1.0\linewidth]{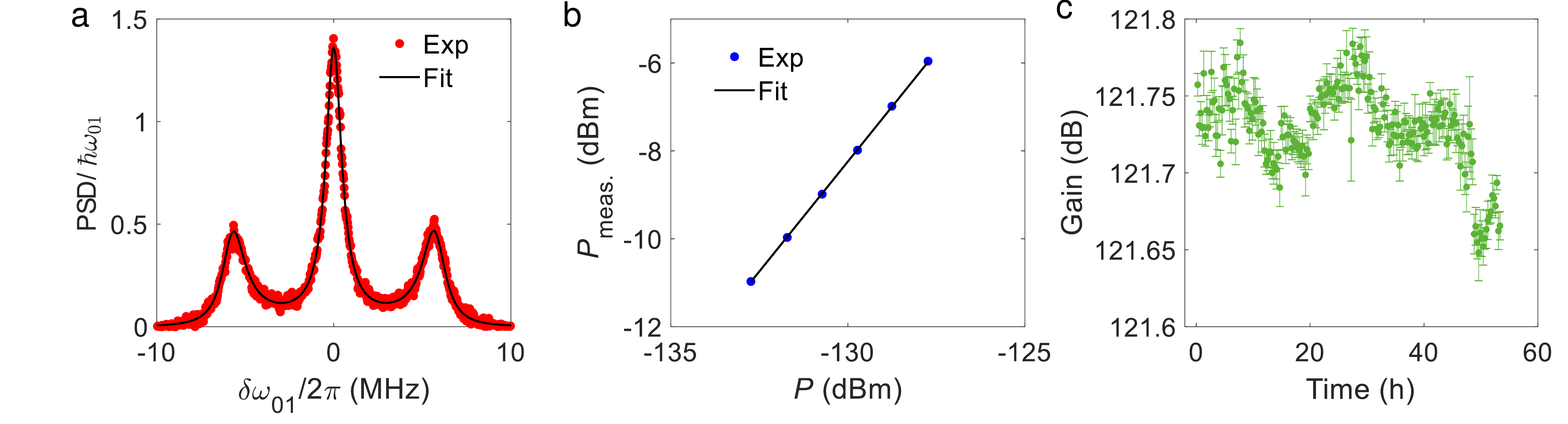}
\caption{\textsf{{\bf{\textsf{{Gain calibration.}}}}
{\bf{a}}, The on-resonant Mollow triplet from qubit emission at $\power=-172.7$ dBm. The solid curve is a fit to extract the Rabi frequency to be $\Omga{} =5.75\pm0.02$ MHz.
{\bf{b}}, Input powers at the qubit ($P$) vs. the measured powers from the output ($P_{\rm{meas.}}$). The black line is the linear fit to obtain the gain in the output line which is $\gain=-121.76\pm0.01$ dB.
{\bf{c}}, Gain fluctuations over 60 hours. The averaged gain is $121.73\pm0.01$ dB.
}
}
\label{gain}
\end{figure*}

Figure.~\ref{gain}(a) shows the power spectrum density (PSD) of the qubit fluorescence at a continuous-wave-pump power, where the reference has been taken into account by subtracting the background. By fitting the data, we obtain the Rabi frequency $\Omga{} =5.75\pm0.02$ MHz. Therefore, the power to the qubit is calculated as $P=10\log_{10}[\hbar\omega_{01}\Omga{}^2/(4\Gama{r})] =-172.7$ dBm. We calibrate other input powers with Mollow triplets and show the results in Fig.~\ref{gain}(b), where $P_{\sf{meas.}}$ presents the measured values from the output line. Therefore, by fitting the data to the equation $P_{\sf{meas.}}=P+G$, we can obtain the gain value $G=-121.76\pm0.01$ dB. We repeat the measurements over 60h  and plot the data with error bars in Fig.~\ref{gain}(c). The mean value of the gain is $121.73$ dB with 0.01dB as the two-standard-deviation error bar. We can see the system gain is very stable. Combined with the input pump powers from the output of the signal generator, we then obtain the attenuation to be 129.7 dB between the signal generator and the sample.

\section{Hamiltonian and Impedance mismatch}\label{sec:sub_impedance}
Our qubit Hamiltonian driven by a continuous wave is ($\hbar=1$)
\begin{equation}
H = -\frac{\Delt}{2}\sigma_z + \frac{\Omga{}}{2}\sigma_x,
\label{A.1}
\end{equation}
where $\omega_{\rm{p}}$ and $\omega_{01}$ are the pump frequency and the qubit $|0\rangle \leftrightarrow |1\rangle$ transition frequency, respectively; $\Delta = \omega_{\rm{p}}-\omega_{01}$ denotes the frequency detuning between the qubit and the pump.

The Lindblad master equation, describing the qubit dynamics with decoherence included, is given by
\begin{equation}
\frac{d}{dt}\rho_q=\mathcal{L}\rho_q = - i[H,\rho_q] + {\mathcal{L}_{\gamma}}\rho_q,
\label{A.2}
\end{equation}
where the Liouvilian $\mathcal{L}_{\gamma}$ is
\begin{equation}
\begin{split}
   \mathcal{L}_\gamma \rho_q& =  \Gama{1}D[\sigma_{-}]\rho_q+\frac{\Gama{p}}{2}D[\sigma_{z}]\rho_q,
\end{split}
\label{A.3}
\end{equation}
in which $D[c]\rho_q= c\rho_q c^\dag - \frac{1}{2}(c^\dag c\rho_q + \rho_q c^\dag c)$ and $\rho_q$ is the density matrix of the qubit state. By solving the master equation, when the qubit reaches its stationary for $t\gg\Gama{1,2}^{-1}$, in the frame rotating at the qubit frequency, we will have
\begin{eqnarray}
\rho_{01,q}&=&\langle \sigma_-\rangle=\frac{\Omga{}\Gama{1}(\Delt - i\Gama{2})}{2(\Omga{}^2\Gama{2} + \Gama{1}(\Delt^2 + \Gama{2}^2))}, \\
\rho_{11,q}&=& \frac{\Omga{}^2\Gama{2}}{2(\Omga{}^2\Gama{2} + \Gama{1}(\Delt^2 + \Gama{2}^2))}.
\label{density}
\end{eqnarray}

{The effect of an impedance mismatch can be understood in a very simplified model as shown in Fig. \ref{impedancemismacthes}.  At the boundary where the impedance mismatch appears, part of the incoming field gets reflected as $r_1a_{\rm{in, o}}$ and the remaining is transmitted as $t_1a_{\rm{in, o}}$. Before reaching the qubit the transmitted field propagates in a different medium where it gains a phase $e^{i\phi_0}$, and its amplitude is attenuated as $\beta$, leading to $t_1a_{\rm{in}}\beta e^{i\phi_0}$ at the qubit position. After interacting with the qubit, according to the input-output theory, we have
\begin{equation}
    a_{\rm out}^{q} = t_1a_{\rm{in, o}}\beta e^{i\phi_0} -i\sqrt{\Gama{r,0}}\sigma_-(t),
\end{equation}
where the corresponding Rabi frequency is $\Omga{}=2\sqrt{\Gama{r,0}}\times t_1a_{\rm{in, o}}\beta e^{i\phi_0}$ with $\Gama{r,0}$ for the capacitive coupling between the qubit and the medium. The field $a_{\rm out}^{q}$ will interfere with the reflected input at the boundary, resulting in
 \begin{equation}
    a_{\rm out}^{\rm{ON}} = (r_1+t_1^2\beta^2e^{i2\phi_0})a_{\rm{in, o}} -i\,t_1\beta\,e^{i\phi_0}\sqrt{\Gama{r,0}}\sigma_-(t),
\label{output}
\end{equation}
where we assume $r_1$ is small enough to ignore the higher orders of reflections.

\begin{figure}
\includegraphics[width=0.5\linewidth]{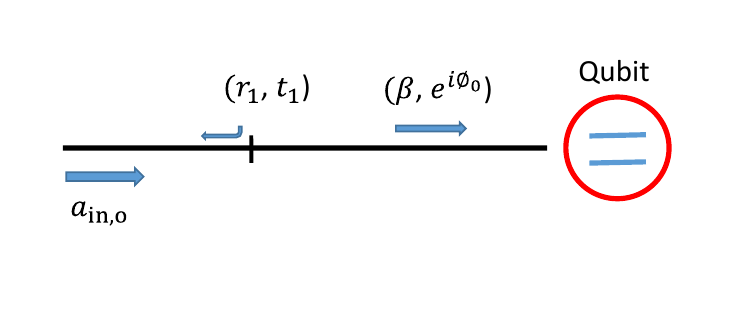}
\caption{\textsf{{\bf{\textsf{{Cartoon of the setup with impedance mismatch.}}}}
$a_{\rm{in,o}}$ is the input signal. $r_1$ and $t_1$ are the corresponding reflection and transmission coefficients at the location where the impedance of the transmission line is changed. The part of the transmission line after the impedance mismatch is taken as a different medium where the propagating wave will obtain a phase $e^{i\phi_0}$ with the amplitude $\beta$ after the attenuation.
}
}
\label{impedancemismacthes}
\end{figure}
When we tune the qubit away, we have
 \begin{equation}
    a_{\rm out}^{\rm{OFF}} = (r_1+t_1^2\beta^2e^{i2\phi_0})a_{\rm{in, o}}.
\end{equation}
By taking the ratio of $a_{\rm{out}}^{\rm{ON}}$ to $a_{\rm{out}}^{\rm{OFF}}$ and replacing $a_{\rm{in}}$ with the Rabi frequency  $\Omga{}$, we have the reflection coefficient as $r=\left\langle a_{\rm{out}}^{\rm{ON}}/a_{\rm{out}}^{\rm{OFF}}\right\rangle$:
\begin{align}
   r &= 1 -\frac{i2\,t_1^2\beta^2\,e^{i2\phi_0}{\Gama{r,0}}}{r_1+t_1^2\beta^2e^{i2\phi_0}}\left \langle \sigma_-(t) \right \rangle, \nonumber \\
   &=1-\frac{i{\Gama{r}}e^{i\phi}{\Gama{1}}(\Delt - i\Gama{2})}{\Omga{}^{2}{\Gama{2}} + \Gama{1}(\Delt^2 + {\it{\Gamma_{\rm{2}}^{\rm{2}}}})}.
\label{reflection}
\end{align}
where we define

 \begin{equation}
    \Gama{r}e^{i\phi} \equiv \frac{\,t_1^2\beta^2\,e^{i2\phi_0}{\Gama{r,0}}}{r_1+t_1^2\beta^2e^{i2\phi_0}},
\label{modifiedGammar}
\end{equation}

\begin{equation}
\phi=\arctan(\frac{r_1\sin2\phi_0}{t_1^2\beta^2+r_1\cos2\phi_0}),
\label{modifiedphi}
\end{equation}
 and
 \begin{equation}
\Gama{r}=\frac{t_1^2\beta^2}{\sqrt{r_1^2+t_1^4\beta^4+2r_1t_1^2\beta^2\cos(\phi_0)}}\Gama{r,0}.
\label{gammar}
\end{equation}
Therefore, we have:
 \begin{equation}
    a_{\rm{out}}^{\rm{ON}}/a_{\rm{out}}^{\rm{OFF}} = 1 -i\frac{\sqrt{\Gama{r}e^{i\phi}}}{a_{\rm{in}}}\sigma_-(t),
\label{output2}
\end{equation}
where $a_{\rm{in}}=\Omga{}/(2\sqrt{\Gama{r}e^{i\phi}})=\sqrt{r_1+t_1^2\beta^2e^{i2\phi_0}}a_{\rm{in,o}}$. Finally, we have the modified input-output equation as:
 \begin{equation}
    a_{\rm{out}} = a_{\rm{in}} -i{\sqrt{\Gama{r}e^{i\phi}}}\sigma_-(t),
\label{output}
\end{equation}
with $a_{\rm{out}}=a_{\rm{out}}^{\rm{ON}}/a_{\rm{out}}^{\rm{OFF}}a_{\rm{in}}=a_{\rm{out}}^{\rm{ON}}/\sqrt{r_1+t_1^2\beta^2e^{i2\phi_0}}$.

Under a weak probe with Rabi frequency $\Omga{pr}\ll\Gama{2}$, Eq.~(\ref{reflection}) reduces to
\begin{equation}
{r=1-\frac{i\Gama{r}e^{i\phi}}{\Delt+i\Gama{2}}.}
\label{weakprobe}
\end{equation}

We can use the equation Eq.~(\ref{modifiedphi}) to estimate where the impedance mismatch appears. We assume that the group velocity is about $0.8948\times10^8$ m/s \cite{wen2019large}. In our setup, both the directional coupler and the SMA (Sub Miniature version A) connector with bonded wires on the sample box could induce impedance mismatches. The distance is 2.5mm between the bonded wires and our qubit, which indicates an added phase $\phi_0\approx0.98$ at the qubit frequency.
We will analyse these two cases below:
\begin{figure}
\includegraphics[width=1\linewidth]{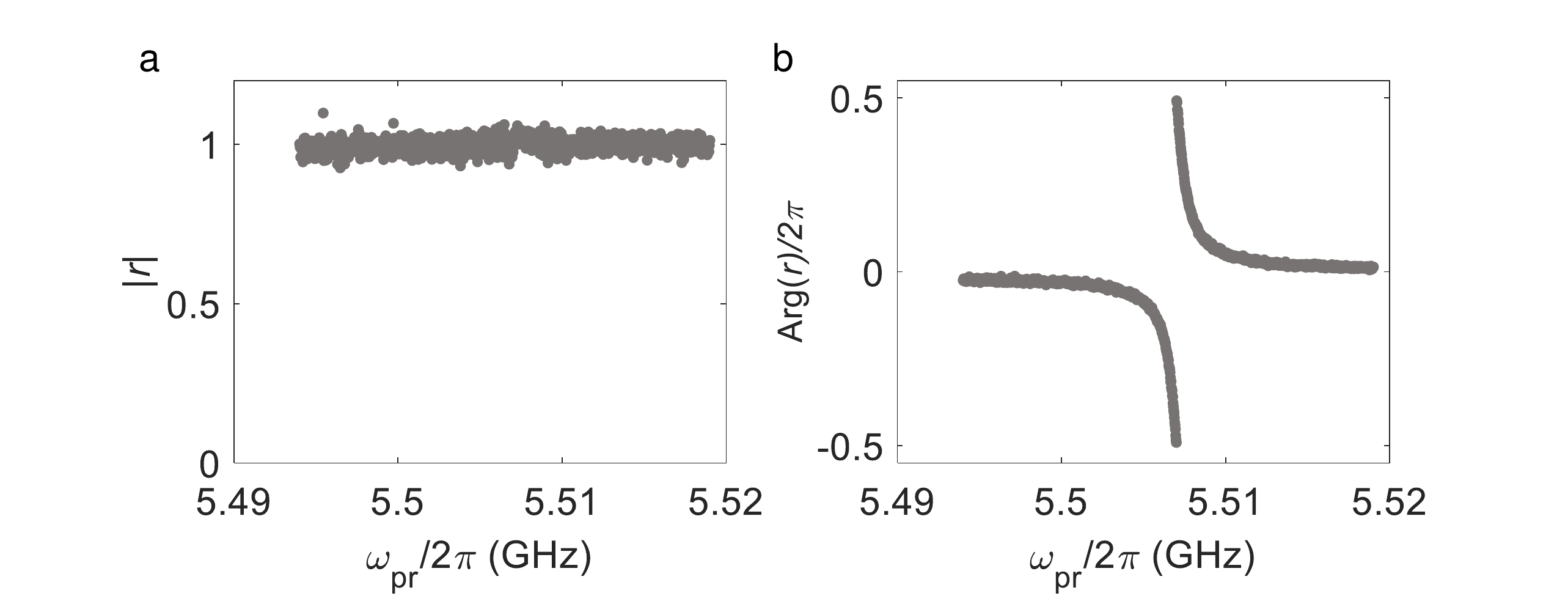}
\caption{\textsf{{\bf{\textsf{Reflection coefficient after compensating the impedance mismatch.}}}}
{\bf{a}}, The magnitude response of the reflection coefficient under a weak probe after compensating the impedance mismatch. The fact that here is no dip at the qubit frequency, implies that the pure dephasing rate and nonradiative decay are negligible for our sample~\cite{lu2019characterizing,scigliuzzo2020primary}.
{\bf{b}}, The corresponding phase response of the reflection coefficient. The raw data is shown in Fig. 1b.
}
\label{setup2}
\end{figure}

(1) For the directional coupler, $r_1=0.1$ in our case, according the extracted $\phi=0.319$, we can get $\beta=\sqrt{0.37\sin{2\phi_0}-0.12\cos{2\phi_0}}\leq0.608$. The value shows that we will have at least 5dB loss between the directional coupler and the qubit, which is very unlikely.

(2) At the SMA connector with bonded wires, it could induce 0.5~dB power loss, corresponding to $\beta\approx0.94$, then we will have $1-5.5956r_1+r_1^2=0$, leading to $r_1\approx 0.18$ which is reasonable. Therefore, we have $\Gama{r}\approx1.07\Gama{r,0}$. So, in our setup, it is very possible that the impedance mismatches are from the SMA connector with bonded wires. Then, we can obtain the compensated reflection coefficient from the measured reflection coefficient as $r_{\rm{com}}=1-(1-r_{\rm{meas.}})e^{-i\phi}/1.07$ as shown in Fig~1(d) in the main manuscript and \ref{setup2}.

From Eq.~(\ref{reflection}), in order to let $r=0$, from $\rm{Im}(r)=0$, we have $\Delt=-\Gama{2}\tan{\phi}$. In addition, from $\rm{Re}(r)=0$, $\Omga{}=\sqrt{\Gama{1}[\Gama{{\rm{r}}}(-\sin{\phi}\Delt/\Gama{2}+\cos{\phi})-\Gama{2}(\Delt^2/\Gamaupdown{2}{2}+1)]}$.
Therefore, $\Omga{}=\sqrt{\Gama{1}[\Gama{{\rm{r}}}/\cos{\phi}-\Gama{2}(\tan^2{\phi}+1)]}$.

For our sample, we have $\phi=0.319$, $\Gama{2}\approx 0.5\Gama{{\rm{r}}}$ and $\Gama{1}\approx \Gama{{\rm{r}}}$, then, $\Omga{}=\sqrt{\Gama{1}[\Gama{{\rm{r}}}(1/\cos{\phi}-\tan^2{\phi}/2)-\Gama{2}]}\approx 0.7061\Gama{\rm{r}}$, which is very close to $0.7071\Gama{\rm{r}}$ as the ideal critical power without any impedance mismatch.

For a resonant probe ($\Delt=0$), Eq.~(\ref{reflection}) is simplified to
\begin{equation}
r = 1-\frac{\Gama{1}\Gama{\rm{r}}e^{i\phi}}{\Omga{}^2 + \Gama{2}\Gama{1}}.
\label{B.3}
\end{equation}

Thus,
\begin{equation}
|r|^2 = 1+\left(\frac{\Gama{1}\Gama{\rm{r}}}{\Omga{}^2 + \Gama{2}\Gama{1}}\right)^2-2\frac{\Gama{1}\Gama{\rm{r}}\cos{\phi}}{\Omga{}^2 + \Gama{2}\Gama{1}}.
\label{B.4}
\end{equation}
When $\Omga{}=\sqrt{\Gama{1}(\frac{\Gama{\rm{r}}}{\cos{\phi}}-\Gama{2})}\approx 0.7431\Gama{\rm{r}}$, the minimal value of  $|r|=|\sin{\phi}|\approx 0.3$ in our case.
We find that the corresponding pump intensity is about $5\%$ higher than the ideal case.
}
\begin{table}

\begin{tabular*}{\columnwidth}{  @{\extracolsep{\fill}} c c c c c c c c  @{} }
  \hline
  \hline
  Sample &$\omega_{01}/2\pi$&$\Gama{r}/2\pi$ & $\Gama{1}/2\pi$ & $\Gama{2}/2\pi$ &$\Gama{n}/2\pi$&$\Gama{p}/2\pi$\\
         &GHz&MHz&MHz&kHz&kHz&kHz\\
  \hline
  \textsf{S1}& 5.50703  & 1.110 & 1.103 (98) & 523 (6)& -10 (100) & -27 (100) \\
  \hline
  \hline
\end{tabular*}
 \caption{\textsf{{\bf{\textsf{Sample parameters}}}. $\Gama{r}$ value is extracted from the reflection coefficient measurement. $\Gama{1}$ and $\Gama{2}$ values are extracted from the decay of the qubit emission. Afterwards, $\Gama{p}$ and $\Gama{n}$ are obtained from $\Gama{p}=\Gama{2}-\Gama{1}/2$ and $\Gama{n}=\Gama{1}-\Gama{r}$, respectively. The numbers in parentheses are error margins of two standard deviations in units of kHz.}} \label{tab:2}
  \centering
  \label{parameters}
  \end{table}

\section{Sample parameters}\label{sec:parameters}
\begin{figure}
\includegraphics[width=1\linewidth]{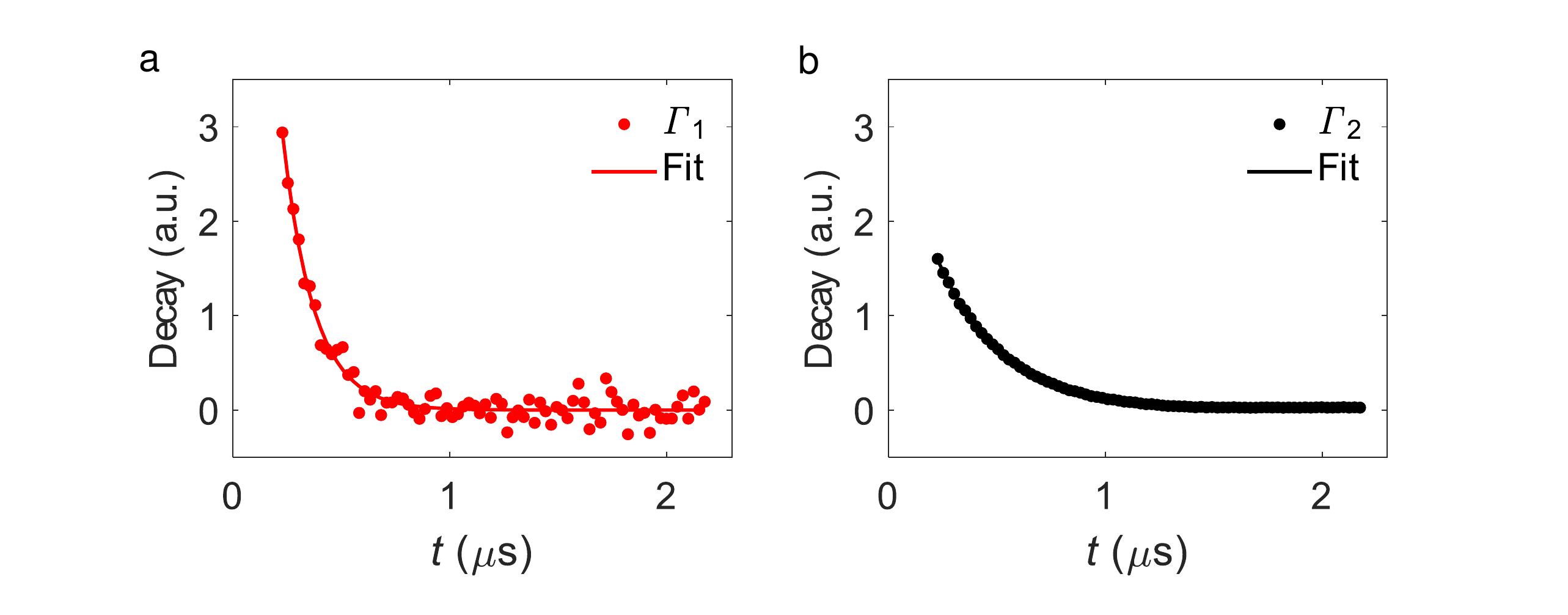}
\caption{\textsf{{\bf{\textsf{Calibrating the atom-field coupling}}.}
{\bf{a}}, This plot shows the power emission (red) of an excited qubit after a short $\pi$-pulse. The corresponding fit (solid curve) indicates  ${\it{\Gamma}}_\textsf{1}/\textsf{2}\pi=\textsf{1.1}\pm0.1$\unite{\textsf{MHz}}.
{\bf{b}}, This plot shows the quadrature decay (black) of an excited qubit after a short $\pi$/2-pulse. We fit the data to extract
${\it{\Gamma}}_\textsf{2}/\textsf{2}\pi=\textsf{0.523}\pm0.006$\unite{\textsf{MHz}}.
}
}
\label{FIGS4_decayrates}
\end{figure}

In this section, we summarize the sample parameters in Table \ref{parameters}. Besides the reflection measurement to extract the radiative decay rate and the the total decoherence rate in the main text, we also send a $\pi/2$-pulse to excite the qubit. By fitting the data from the quadrature decay and the power decay of the qubit shown in Fig.\ref{FIGS4_decayrates}, we obtain the total relaxation rate $\Gama{1}/2\pi=1.1\pm0.1\unite{MHz}$ and the decoherence rate $\Gama{2}/2\pi=0.523\pm0.006\unite{MHz}$.

\section{Steady-state purity, coherence and population}\label{sec:purity}
From Eq.~\ref{density} we have the steady-state qubit density matrix
\begin{gather}
\rho_q=
 \begin{bmatrix} \rho_{00,q}& \rho_{01,q} \\ \rho_{10,q} & \rho_{11,q} \end{bmatrix}
 =\begin{bmatrix}
   1-\frac{\Omga{}^2}{2(\Omga{}^2+\Gama{1}\Gama{2})}&
   \frac{-i\Omga{}\Gama{1}}{2(\Omga{}^2+\Gama{1}\Gama{2})} \\
   \frac{i\Omga{}\Gama{1}}{2(\Omga{}^2+\Gama{1}\Gama{2})} &
   \frac{\Omga{}^2}{2(\Omga{}^2+\Gama{1}\Gama{2})}
   \end{bmatrix}.
\end{gather}

\begin{figure}
\includegraphics[width=0.5\linewidth]{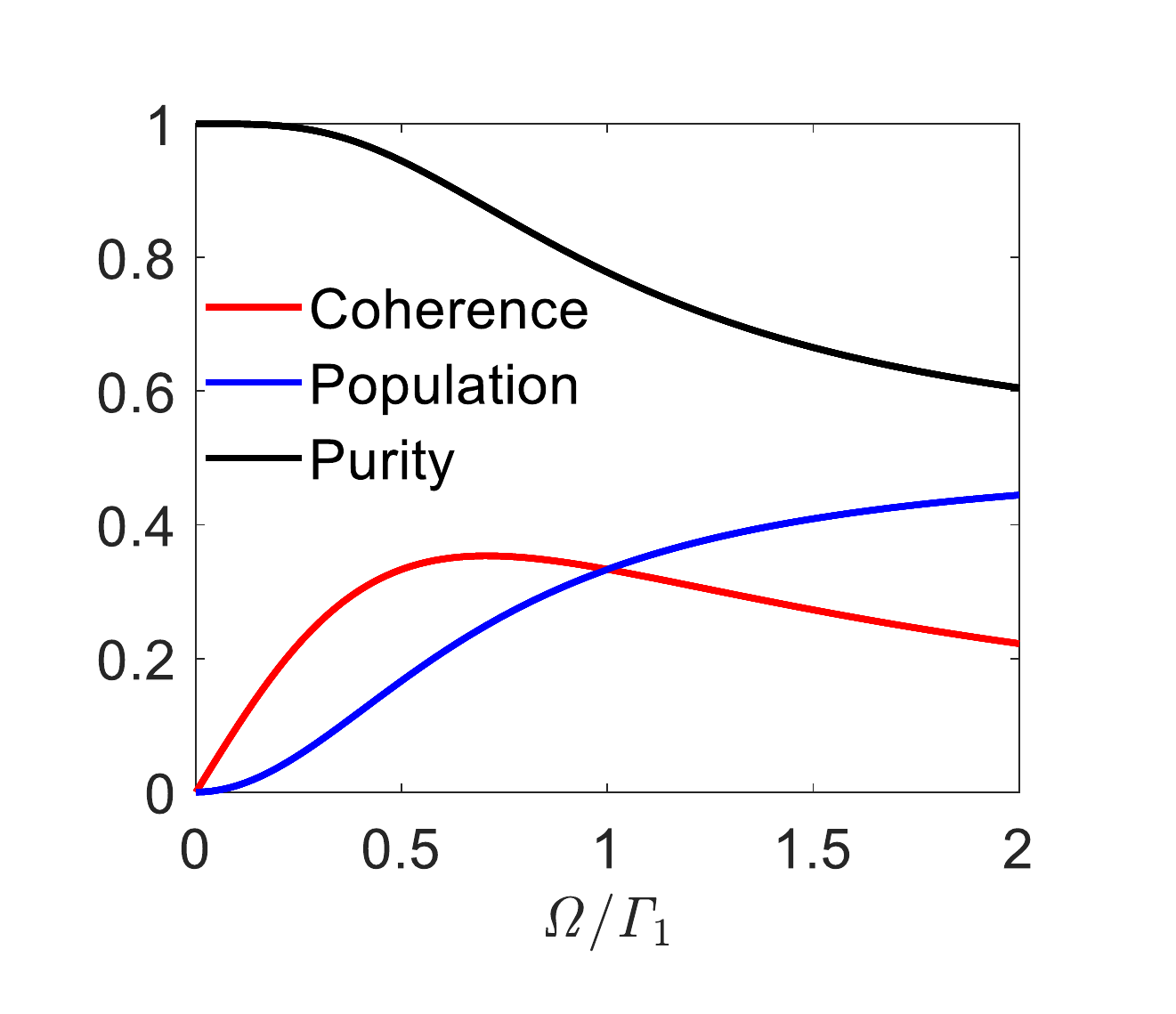}
\caption{\textsf{{\bf{\textsf{{Theoretical coherence, population and purity of the qubit state.}}}}
Qubit steady-state coherence, excited state population and purity as a function of the Rabi frequency of the drive.}
}
\label{Qubitstate}
\end{figure}

The coherence of the qubit state is $|\rho_{01,q}|$, and the purity corresponds to $Tr(\rho_q^2)=\rho_{00,q}^2+\rho_{11,q}^2+2|\rho_{01,q}|^2$. By taking ${\partial |\rho_{01,q}|}/{\partial\Omga{}}=0$, we have the maximum value, $\sqrt{\Gama{1}/(16\Gama{2})}$ of the coherence at $\Omga{}=\sqrt{\Gama{1}\Gama{2}}$ which equals to the critical power when the qubit has zero non-radiative decay rate. $\rho_{00,q}$ and $\rho_{11,q}$ are the corresponding populations of the qubit ground and excited states, respectively. In Fig. \ref{Qubitstate}, we calculate these quantities with different Rabi frequencies.

\section{Qubit response at the critical power}\label{sec:magic}
\begin{figure*}[tbph]
\includegraphics[width=1\linewidth]{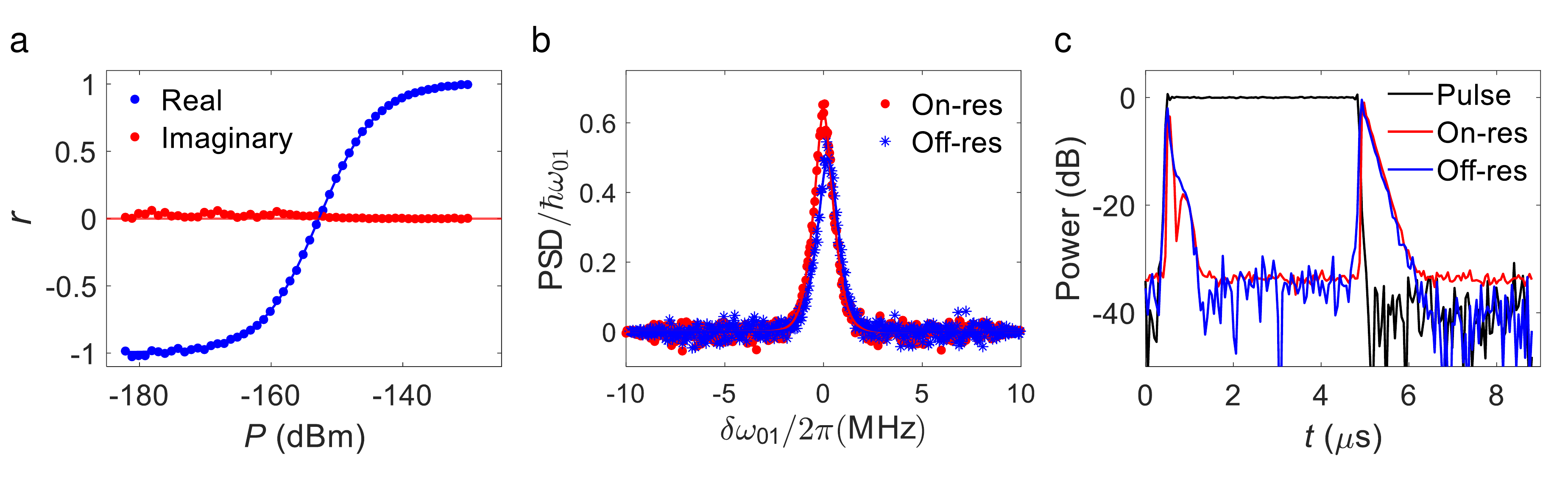}
\caption{\textsf{{\bf{\textsf{Qubit response at the critical power}}.}
{\bf{a}}, The plot shows the reflection coefficient, \textsf{\textit{r}} as a function of off-resonant incident powers, $P$. The real and imaginary responses are shown in blue and red, respectively.
{\bf{b}}, Power spectrum density (PSD) of the outcoming field. Red and blue markers are for the on/off resonant cases, respectively. In panel {\bf{a}} and {\bf{b}}, solid curves are the corresponding fits.
{\bf{c}}, Reflected pulses at the critical power with time \textsf{\textit{t}}. The black curve is measured when the qubit is tuned away from the pulse frequency and its magnitude is normalized to unity. Red and blue curves are the responses for the on/off resonant pulses, respectively.
}
}
\label{magicpower}
\end{figure*}
In Fig.~1b we measure the reflection coefficient around the critical power (orange dots) as a function of the pump frequency and show the complex data in a circle plot. Again, due to the part of the incident photons reflected from the place where the impedance mismatch appears, the circle is rotated leading to a nonzero value at the qubit frequency shown as a big red dot where the pump frequency is on resonance. However, we also find an off-resonance frequency marked by the big blue dot where $a_{\rm{out}}=0$. This corresponds to a frequency
$170\unite{kHz}$ larger than the qubit frequency. This as the coherent photons reflected by the mirror and the qubit interfere destructively with the reflected photons at this detuned frequency due to the impedance mismatch. To find the critical power more precisely at the big blue dot, in Fig.~\ref{magicpower}a, we plot $r$ as a function of the pump power $P$ where the accurate critical power is $P_{\rm{m}}=-152.8\unite{dBm}$.  In the figure, the real part shows an added $\pi$ phase shift due to the reflection by the qubit when $P$ is weak, in contrast to when $P$ is strong enough to saturate the qubit.
We also measure the power spectrum of the output emission for a qubit driven at the critical power, shown in Fig~\ref{magicpower}b. By fitting the data to the power spectrum density equation \cite{lu2019characterizing}, we extract $\Omga{m}= (0.74\pm0.01)\Gama{1}$ and $\Omga{m}= (0.72\pm0.02)\Gama{1}$ for the on- and off-resonance cases respectively. These values are close to the ideal case $\Omga{m}= 0.707\Gama{1}$ without any impedance mismatch.

Having characterized our sample and the critical power, the next step is to measure the Wigner function of the output field in a wave packet mode.
In order to extract the Wigner function from the measured signal, we need to get the background as the reference. We generate a pulse at the critical power and send it to the input port in Fig.~1a. Thus, the background can be obtained by measuring the system noise when the pulse is off. The pulse power is calibrated at the critical power, and it is normalized and shown in black in Fig.~\ref{magicpower}c. The pulse length is $4.4~\mu$s, much longer than the lifetime of the qubit, $T_1=1/\Gama{1}=145\unite{ns}$ . This is in order to ensure that the qubit has reached the steady-state before we perform our measurement of its output field.
{The red and blue curves correspond to the output emission when the qubit is driven
on- and off-resonance respectively (red and blue dots in Fig. 1b). For the on-resonance case, in order to achieve the cancellation of the coherent output, i.e.,  $\langle a_{\rm{out}} \rangle = 0$, we have used a second pulse with a $\pi$-phase difference compared to the input pulse and fed it into the cancellation port in the setup in Fig.~1a}. For both the on- and off-resonance cases, the qubit evolves into the steady state after 1.5 $\mu s$ and the coherent signal is suppressed by up to $-34\unite{dB}$.  At the end of this pulse the qubit decays back to the ground state, leaving the background noise.

\begin{figure*}[tbph]
\includegraphics[width=1\linewidth]{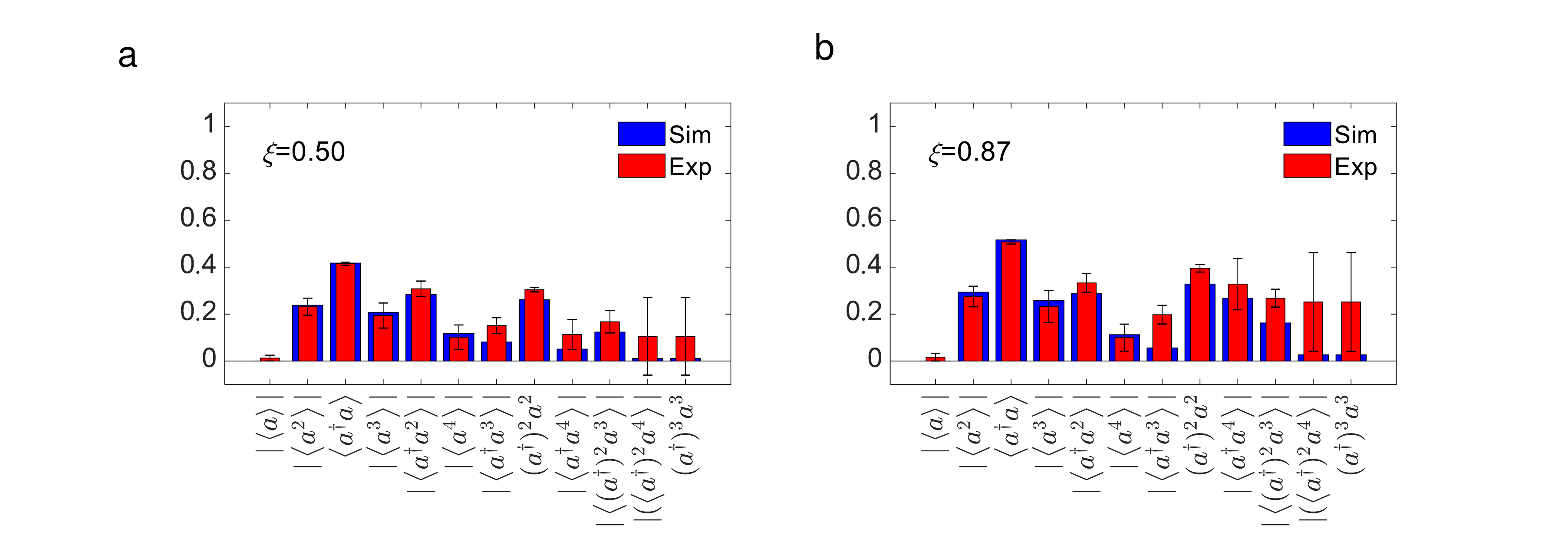}
\caption{\textsf{{\bf{\textsf{Moments for a Gaussian filter}}.}
Different orders of moments for the propagating state from the qubit emission in the mode defined by a Gaussian filter of the standard deviation $\xi$. The red rectangles correspond to experimental data for the off-resonant case. The blue rectangles correspond to the numerical simulation of the ideal line, without the impedance mismatch. The error bars are the corresponding standard deviations for the extracted moments. In both panels, the low-order moments are consistent with the results from the numerical simulation. However, due to the system noise, the high-order moments have a relatively larger error bar. Compared {\bf{b}} to {\bf{a}}, with a higher photon population, the high order of moments becomes non-negligible. Therefore, moments up to only sixth order do not contain all the information, leading to the disagreement of the high-order moments between the simulation and the experiment in {\bf{b}}.
}
}
\label{momentgaussian}
\end{figure*}

\begin{figure*}[tbph]
\includegraphics[width=1\linewidth]{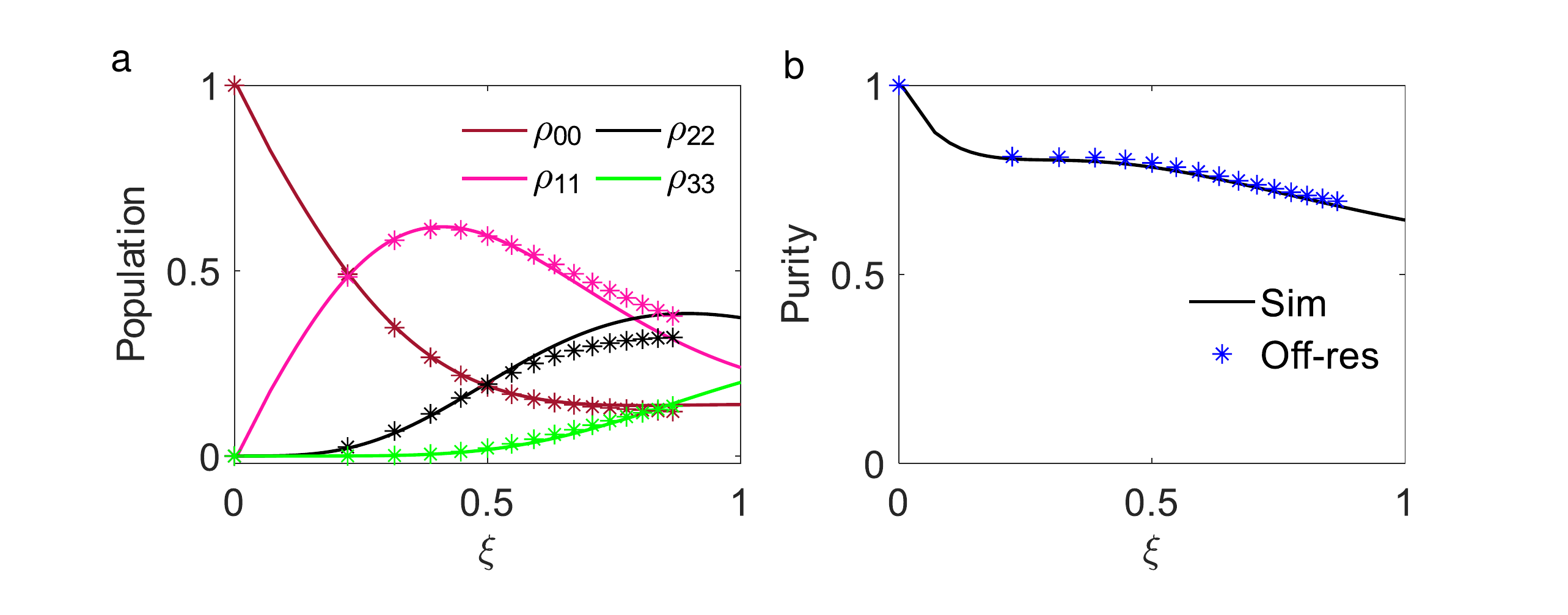}
\caption{\textsf{{\bf{\textsf{Population and purity with a Gaussian filter as a function of $\xi$}}.}
$\xi$ is the standard deviation of the Gaussian filter.
{\bf{\textsf{a}}}, The photon population from the qubit emission.
{\bf{\textsf{b}}}, The purity of the total emission field.
In both panels, the markers are from experiments and the solid curves are from the numerical simulations. The experimental results agree well with the theory.
}
}
\label{wignergaussian}
\end{figure*}

\section{Moment, population and purity with a Gaussian filter}\label{sec:gaussian}
\begin{figure*}[tbph]
\includegraphics[width=1\linewidth]{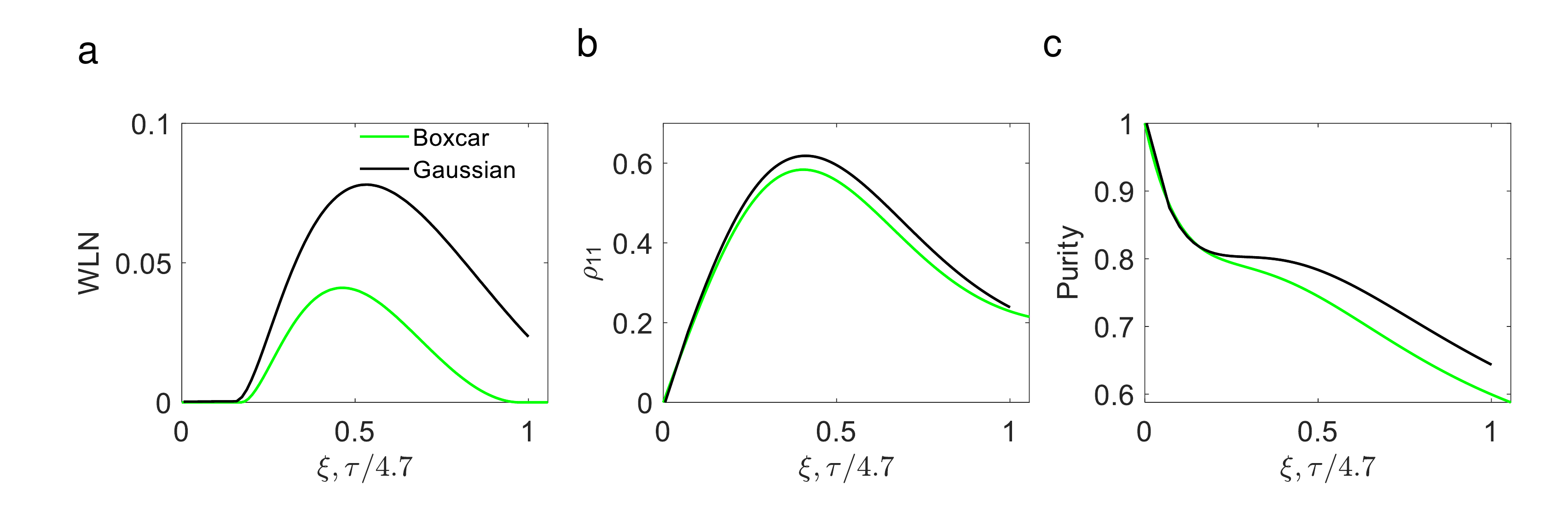}
\caption{\textsf{{\bf{\textsf{Numerical comparison between a Gaussian and a boxcar filter}}.}}
{\bf{\textsf{a}}}, $\sf{WLN}$,
{\bf{\textsf{b}}}, Single-photon population from the qubit emission,
{\bf{\textsf{c}}}, and purity of the emission field for a Gaussian filter and a boxcar filter as a function of the standard deviation of the Gaussian filter and the rescaled width of the boxcar filter.
}
\label{filtercomparison}
\end{figure*}

Figure~\ref{momentgaussian} a and b show the extracted different order of moments from the qubit emission after a Guassian filter with $\xi=0.5$ and $\xi=0.87$, respectively.
Figure~\ref{wignergaussian} a shows the photon population from the qubit emission after the Gaussian filter, which is obtained by applying a displacement, $2^{3/4}\pi^{1/4}\sqrt{\xi}*\Omga{m}/2\sqrt{\Gama{r}\Gama{2}}$, onto the total output emission. The corresponding purity of the emission field is shown in Fig.~\ref{wignergaussian} b. To understand why the Gaussian filter can give a larger Wigner negativity compared to a boxcar, we plot the corresponding single-photon population and the purity from the numerical simulation together, as shown in Fig.~\ref{filtercomparison}. The time axis for the boxcar is re-scaled. Clearly, we find that by using a Gaussian filter, both the single-photon population and the purity are higher than the boxcar filter, resulting in a higher value of $\sf{WLN}$.

\begin{figure*}
\includegraphics[width=1\linewidth]{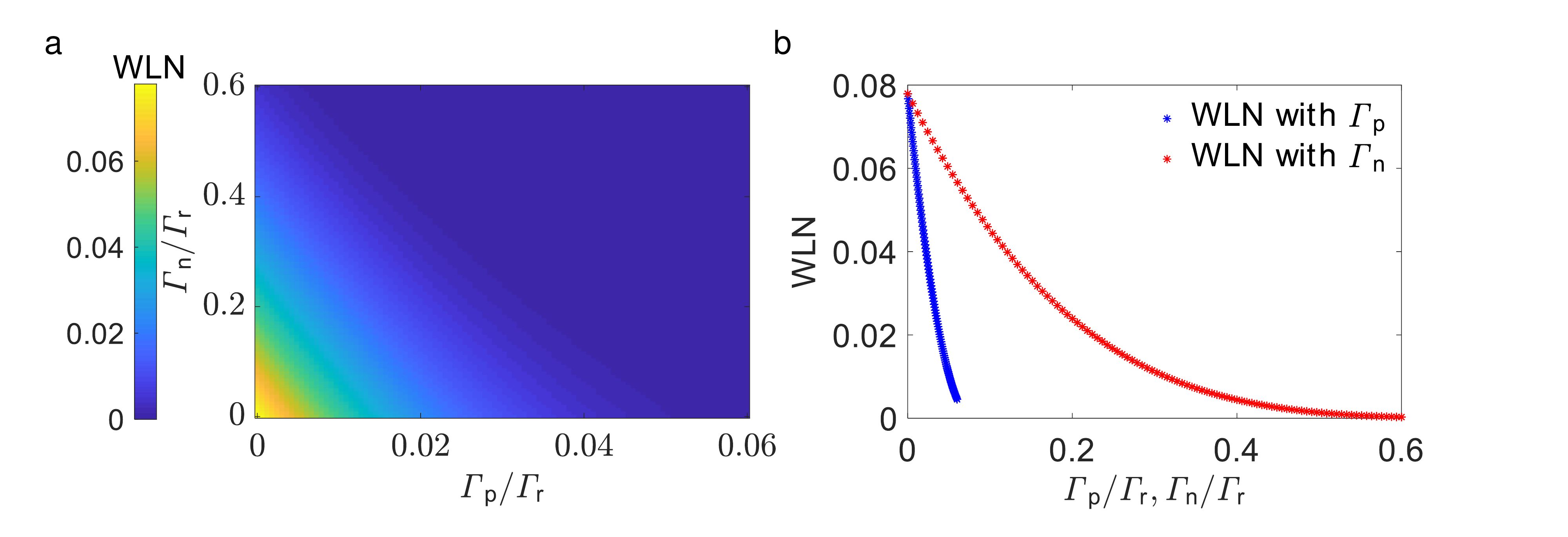}
\caption{\textsf{{\bf{\textsf{{Estimation for the frequency tunable range for our non-classical source.}}}}
{\bf{a}}, Numerical maximum WLN with the pure dephasing rate $\Gama{p}$ and the non-radiative decay $\Gama{n}$.
{\bf{b}}, WLN vs. different $\Gama{p}$ values with $\Gama{n}=0$ or different $\Gama{n}$ values with $\Gama{p}=0$ in {\bf{a}}.
}
}
\label{wln-dephasing}
\end{figure*}

\section{Frequency tunability of the non-classical source}\label{sec:dephasing}
Even though our sample has a minuscule pure dephasing rate and non-radiative decay rate at the flux sweet spot, we also numerically study the influence from both of these loss channels on $\sf{WLN}$ values with Gaussian filters. It will help us to estimate the frequency tunable range of our source because our sample has a SQUID loop which allows us to tune the qubit frequency through the external flux. Plenty of studies demonstrate that the non-radiative decay rate is basically limited by the two-level fluctuators, which can be quantified as the loss tangent \cite{martinis2005decoherence,burnett2014evidence,muller2019towards}. The loss tangent factor decreases when the qubit frequency is tuned down. Therefore, we can ignore the effect from non-radiative decay for our sample since this value is quite small at the maximum qubit frequency as shown in Fig.~\ref{setup2}. However, the pure dephasing rate will be increased when the qubit frequency is tuned down by the external flux.

The maximum WLN values in Fig.~\ref{wln-dephasing} a are obtained by optimizing the variance, $\xi$, of the Gaussian filter. We find that both the pure dephasing rate and nonradiative decay rate can decrease the negativity significantly. Notably, the negativity is much more sensitive to the pure dephasing rate compared to the nonradiative decay rate. In Fig.~\ref{wln-dephasing} b, without the pure dephasing rate, even though the non-radiative decay rate is up to 0.023$\Gama{r}$, corresponding to 25$\unite{kHz}$, the WLN value can be still around 0.07. However, with the pure dephasing rate $\Gama{p}/2\pi=25\unite{kHz}$ and $\Gama{n}=0$, the WLN decreases to be half of the maximum WLN. This becauses the purity of the state decreases faster with pure dephasing than with nonradiative decay. Here, we use the half value of the maximum WLN to investigate the possible frequency-tunable range of our source. From Ref~\cite{yong2020single}, we measured the flux noise in the environment, which is a $1/f$ type noise with spectral density $S_{\Phi}(f)=A_{\Phi}/f$. The value of $A_{\Phi}^{1/2}=2\unite{\mu\Phi_0}$ is close to others in Ref. ~\cite{hutchings2017tunable}. This type of noise give us a pure dephasing rate with the relationship $\Gama{p}=\sqrt{A_{\Phi}|\ln(2\pi f_{\rm{IR}}t)|}\frac{\partial \omega_{01}}{\partial\Phi}$ \cite{hutchings2017tunable,lu2021quantum}. Then, we estimate the pure dephasing rate to be around 25\unite{kHz} at $\omega_{01}/2\pi=5.1\unite{GHz}$. Thus, the possible frequency-tunable range of our non-classical source can be up to 400\unite{MHz}. This range could be enhanced further by engineering the radiative decay of the qubit even larger.

\bibliographystyle{naturemag}